\documentclass[12pt,aip,jcp,amssymb,amsmath,superscriptaddress,showpacs,floatfix,reprint]{revtex4-2}

\usepackage{mathtools}
\usepackage{epsfig}
\usepackage{graphicx}
\graphicspath{{figures/}}
\usepackage[caption=false]{subfig}
\usepackage[english]{babel}
\usepackage{xcolor} 
\usepackage{soul}
\usepackage{dcolumn}
\usepackage{bm}

\usepackage[utf8]{inputenc}
\usepackage[T1]{fontenc}
\usepackage{etoolbox}
\usepackage[unicode=true]{hyperref}
\hypersetup{breaklinks=true, colorlinks=true, urlcolor= blue, citecolor=blue}
\newcommand{\ie}{\emph{i.e.}}

\makeatletter
\def\@email#1#2{%
	\endgroup
	\patchcmd{\titleblock@produce}
	{\frontmatter@RRAPformat}
	{\frontmatter@RRAPformat{\produce@RRAP{*#1\href{mailto:#2}{#2}}}\frontmatter@RRAPformat}
	{}{}
}%
\makeatother
\begin{document}
	
	\preprint{AIP/123-QED}
	
	\title[Wetting in off-critical mixtures]{Early-time wetting kinetics in surface-directed spinodal decomposition for off-critical quenches: A molecular dynamics study }  
\author{Syed Shuja Hasan Zaidi}%
\affiliation{Department of Physics, Indian Institute of Technology Jodhpur, Jodhpur 342030, India}
\author{Saumya Suvarna}
\affiliation{Department of Physics, Birla Institute of Technology Mesra, Ranchi  835215, India}
\author{Madhu Priya}
\affiliation{Department of Physics, Birla Institute of Technology Mesra, Ranchi  835215, India}
	\author{Sanjay Puri}
\affiliation{School of Physical Sciences, Jawaharlal Nehru University, New Delhi 110067, India}
\author{Prabhat K. Jaiswal}

\affiliation{Department of Physics, Indian Institute of Technology Jodhpur, Jodhpur 342030, India}
\email[Corresponding author:\;]{prabhat.jaiswal@iitj.ac.in}
	\date{\today}
	
	\begin{abstract}
		We present results from the molecular dynamics (MD) simulation of surface-directed spinodal decomposition (SDSD) in binary fluid mixtures ($A+B$) with off-critical compositions. The aim is to elucidate the role of composition ratio in the early-time wetting kinetics under the influence of long-range surface potential. In our simulations, the attractive part of surface potential varies as $V(z)= -\epsilon_a/z^{n}$, with $\epsilon_{a}$ being the surface-potential strength. The surface prefers `$A$' species to form the wetting layer. Its thickness [$R_1(t)$] for the majority wetting (number of $A$-type particles [$N_A$] > number of $B$-type particles [$N_B$]), grows as a power-law with an exponent $1/(n+2)$. This is consistent with the early-time kinetics in the form of potential-dependent growth present in the Puri-Binder model. However, for minority wetting ($N_A$ < $N_B$), the growth exponent in $R_1(t)$ is less than $1/(n+2)$. Furthermore, on decreasing the field strength $\epsilon_{a}$, we recover $1/(n+2)$ for a minority wetting case. We provide phenomenological arguments to explain the early-time wetting kinetics for both cases. 
	\end{abstract}
	
	\maketitle
	
	\begin{quotation}
		
	\end{quotation}
	
	\section{\label{sec:level1}INTRODUCTION}
The kinetics of phase separation have attracted immense scientific interest and have elucidated the roles played by the physical properties of the binary fluid mixtures $(A+B)$. For instance, the dynamics associated with the viscous and inertial motion of the fluid may alter the growth of the phase-separating domains.\cite{VS09,tanaka01,bray02,onuki02} It is crucial to understand the domain growth process and the laws related to its coarsening, \cite{VS09,bray02,onuki02} to control the morphological patterns that, when frozen, will produce a material with specific properties. The tremendous application of this demixing process in technologies like plastic electronics and membrane fabrications, \cite{KYM19,WYG19,JDM15,OB12} alloys and glasses, \cite{DEE14,MH21} and complex fluids like polymer blends, \cite{MRA95,RLE91} emulsions, \cite{MJ14,SD11} and ionic liquids (ILs) \cite{JFJ09} have paved the way in designing materials with enhanced mechanical stability \cite{MJH09,TH16,KY19,MP20,PV14,ZLY21} and stabilization of proteins. \cite{YNH20} Further from a theoretical perspective, phase-separation kinetics invokes concepts of pattern formation (and its prediction via machine learning \cite{CMY23}), self-similarity, finite-size scaling, etc. \cite{bray02,gunton83phase} 

\emph{Spontaneous decomposition} of a binary mixture happens when the homogeneously mixed system at a high temperature turns thermodynamically unstable following a sudden quench well inside the spinodal curve. Subsequently, the components segregate with the domain formation of $A-$rich and $B-$rich phases, demarcated by interfacial boundaries. Then, the system proceeds towards a new equilibrium with the growth in average domain size, $l(t)$, of either phase by minimizing interfacial contributions to the relevant thermodynamic free energy. In the scattering experiments, the scattering peak corresponds to $k_m=2\pi / \lambda_c$, with a characteristic length scale, $\lambda_c \simeq l(t)$. The peak shifts toward lower wavevectors following the decreasing number of interfaces as the domains grow. Thus, the domain coarsening can be characterized by tracking the temporal evolution of $l(t)$ captured by scattering experiments. \cite{VS09,bray02,onuki02} For binary mixtures, the coarsening follows a power law, $l(t) \sim t^{\theta}$, and the reported values of the exponent ($\theta$) articulate the modes of transportation involved. The prominent mode in binary alloys is evaporation-condensation  \cite{IV61} or Brownian coagulation, \cite{siggia79,ohta84ap} which involves molecular transportation with $\theta = 1/3$.  For binary fluids, surface-tension driven flows\cite{siggia79} followed by collective motion due to droplet's inertia\cite{furukawa85} results in mass transport, producing $\theta \sim 1$ and $\theta \sim 2/3$, respectively.     
	
Ignoring the boundaries or surface effects simplifies the phase separation of an unstable mixture, as the system exhibits isotropically growing fluctuations. But it is seldom true that the demixing system is free from such inhomogeneities. Especially, in thin films used for plastic electronics, where fabrication of active medium using binary mixtures is done. During the fabrication process, the film morphology is affected due to interacting surfaces and the film's finite size. \cite{JDM15,SMJ97,Michels11} Furthermore, the morphological changes affect the percolation pathways for the charges created, which is a crucial parameter for the operation of active mediums in the devices. Surface effects are prevalent in processes where the demixing happens in a container, and its walls prefer to attract a specific species in the binary mixture. In such cases, the wetting happens and the preferred species coats the surface with a semi-macroscopic wetting layer while demixing happens spontaneously. This phenomenon is termed \emph{surface-directed spinodal decomposition} (SDSD). \cite{puri05,KSS10} The SDSD is inevitable and has great technological relevance in fabricating low-cost plastic electronics, organic photovoltaics (OPVs), \cite{LEV05,JRI10,KDB08,Chen19,YDL11,Michels11,HYX20,CJP17} organic thin films, \cite{PMA21} etc. In particular, tailoring the domain morphologies and freezing them has proved beneficial for enhanced charge collection in OPVs, \cite{Chen19,YDL11,Michels11,HYX20,CJP17} tuning physical, mechanical, and surface properties of polymer blends,  \cite{MRA95,YDL11,MPJ16} as electrolytes for energy storage, \cite{TG19} in bit-patterned media,  \cite{CTJ16} nanowires, \cite{PAC15} polarizers, \cite{SJI14} ion conduction channels, \cite{JGH18} nanolithography,  \cite{CMD14} and many more applications. \cite{PMA21,MAM21,ERD12} 	

The SDSD breaks the symmetry of the demixing system perpendicular to the surface and introduces additional length scales characteristic of the phase ordering normal and parallel to the surface, producing anisotropic domains. \cite{puri05,KSS10} Many scientific measures have been taken up using experiments, \cite{MRA95,RLE91,MG03,krausch95, PA91,APF92,BCA93,Michels11,YDL11} analytical investigations, \cite{SK92,troian93prl,troian92,HPW97,RR90,KH91,marko93} and computer simulations to investigate the SDSD process. \cite{SY88,SH97,puri05,binder83,GA92,SK01,SK94,PSS12,SSJ01,SSJ05,SSJ06-pre,bray02,APF21,toxvaerd99prl,HA97pre,PWA94prl,HT00,tanaka01,SAR20,PPS20,PPS20-2,MPJ16} The Puri-Binder (PB) model of SDSD \cite{SK92} is the first successful numerical model. The model employs the master-equation approach to derive a coarse-grained Cahn-Hilliard-Cook (CHC) equation for the evolution of the order-parameter (concentration difference) at a spatial point $\vec{r}$.  
The model is also supplemented with two appropriate boundary conditions for surface contributions to the evolution. They studied domain coarsening for critical mixtures ($50\%A - 50\%B$) with short-ranged $ \delta $-function potential for the surface and found the logarithmic growth of the wetting layer thickness, $R_1(t)$, when the surface is completely wet by the preferred component. They characterized the phase ordering parallel and perpendicular to the surface and analyzed the temporal evolution of the characteristic length scale, $l_{||} (t)$, and $l_\perp (t)$, respectively, in surface vicinity and bulk. 

Puri and Binder \cite{SK94,SK01,SK02pre} also simulated the effects of the long-range wall potential of the form $ V(z) \sim z^{-n} $ for complete wetting cases. They found potential-dependent growth $R_1 (t) \sim t^{1/(n+2)}$ at early times for critical \cite{SK94} and off-critical \cite{SK01,SK02pre} quenches. Later, the potential dependent regime makes the crossover to a diffusive regime. However, this crossover was not observed for the majority wetting, where the preferred species (wetting component) is in the majority. For the majority wetting cases, they claimed that the potential-dependent regime is present for arbitrary times, and no crossover to diffusive growth is observed. The rationale behind it is the negative curvature of the minority droplets formed in bulk, which creates a negative current of wetting components toward them. Although these results are important, they are limited to the case of diffusive transport. It is physically relevant to theoretically ascertain the role of advective modes due to hydrodynamics. \cite{SK01,SK02pre} 

There are results from hydrodynamics-enabled studies where fluid velocity has proven to drastically and substantially alter the domain growth (wetting layer and bulk domains) for critical and near-critical quenches.\cite{ERD11jpcb} Jamie \emph{et al.} conducted experiments on a colloidal-polymer solution and showed how at late times, the tubes of the wetting component from the bulk perforate through the depletion layer and connects with the wetting layer, overall accelerating the coarsening process. Such acceleration is similar to Siggia's mechanism\cite{siggia79} of surface-tension-driven flows in bulk and thus produces a power-law growth exponent of $1$. Conversely, they proved that in the absence of such perforations by tubes for near-critical mixtures with minority wetting (wetting component is in minority), the wetting layer growth remains diffusive, signifying the role played by the connectivity of bulk and wetting layer by the tubes of hydrodynamical length scales. Along similar lines but with majority wetting, Krausch, \cite{GEF94} in his experiment, did not notice any retardation of the wetting layer due to connectivity maintained between the bulk and the wetting layer, allowing the wetting component to flow around the minority droplets and reach the wetting layer. Moreover, they also performed diffusion-limited simulations and found no acceleration whatsoever, contrary to experiments that incorporate hydrodynamics naturally. In another hydrodynamics-enabled numerical study by Chen and Chakrabarti, \cite{HA97pre} the authors found that after switching off the hydrodynamics, the wetting layer grew as $\sim t^{1/3}$ for some time and slowed down later, citing similar negative currents seen in the diffusive PB models. On the other hand, retardation was absent in the presence of hydrodynamics. Thus it is now a fact that the fluid velocity accelerates the wetting layer growth when the majority phase wets the surface. \cite{ERD11jpcb,GEF94,HA97pre,tanaka01}

Most of the studies we discussed primarily focused on the asymptotic time and length scales and say nothing about how early-time coarsening of domains is affected by the hydrodynamic modes, if present. Therefore, the present work uses molecular dynamics (MD) to investigate the early-time dynamics associated with the wetting layer growth for off-critical mixtures. We examined two kinds of off-critical mixtures ($N_A\ne N_B$, where $N_A$ and $N_B$ is the number of $A$-type and $B$-type particles, respectively) corresponding to majority wetting ($N_A > N_B$) and for minority wetting ($N_A<N_B$). The wetting component ($A$) remains the same in both cases. The model and methodology are the same as in our previous work \cite{SPMS22pre} and the other previous studies \cite{PSS10, PSS12, PSS12-epl, SPMS22pre} done investigating wetting layer growth laws in critical mixtures. We have mainly focused on the early-time regimes for power-law wall potential. The features of our resultant nonequilibrium simulations resonate with the works by Keblinski \emph{et al}. \cite{PWA94prl, PWA93} Our simulations show molecular layering near the surface and the rapid creation of the wetting layer for the majority wetting. Our efforts at present are to establish validations for the growth exponents during the early wetting layer growth and to elucidate the role of hydrodynamics at these lengths and time scales. What we found is the disparity in the wetting layer growth for very early-time regimes across the compositions, which has not been established elsewhere. For us, the early-time regimes comprise potential-dependent growth for the majority wetting. However, for minority wetting, the wetting layer growth does not agree with the theoretical prediction from the PB model, instead, it shows slower growth than expected. The departure is seen for different ranges of power-law potential and the degree of off-criticality. Furthermore, the potential-dependent growth is recovered for minority wetting when the field's strength of the long-range attractive potential is reduced. The depletion layers formed for minority wetting cases have compositions from the vicinity of the nucleation regime in the phase diagram and create a local barrier for the flux of wetting particles toward the wetting layer. Note that an opposite effect has been seen in earlier studies\cite{GEF94,GAJ94pre,HA97pre} of majority wetting with a global composition of highly-off critical value. For these cases, the expulsion of the non-wetting minority phase from the vicinity of the surface may lower the local nucleation barrier in bulk, leading to the formation of minority phase domains near the surface. However, we could not see such structure formations in the depletion layer as our simulations are restricted to very early times with negligible phase separation.

	\section{\label{sec:level2}SIMULATIONS}
	\subsection{Molecular Dynamics (with Hydrodynamics)}
To carry out the study on SDSD, we perform molecular dynamics simulation using LAMMPS software. \cite{plimpton95} The model is similar to that employed in our earlier studies of critical mixtures on flat surfaces. \cite{SPMS22pre,PSS10,PSS12,PSS12-epl} We begin with a binary mixture $(A+B)$ in a simulation box of a volume $V = L\times L \times D$. The box maintains semi-infinite geometry by being periodic in $ x, y $ directions, and holds confining walls in the $z$ direction. We model the wall-particle interaction by an integrated Lennard-Jones (LJ) potential of the form as given below: 
\begin{equation}\label{eq1:wall}
u_w(z)=\frac{2\pi \rho_N \sigma^3}{3}\left[ \frac{2\epsilon_r}{15}\left( \frac{\sigma}{z^\prime}\right)^9 - \delta_\alpha \epsilon_a\left(\dfrac{\sigma}{z^\prime}\right)^n \right].
\end{equation}
The parameters $\epsilon_r$ and $\epsilon_a$ correspond to the repulsive and attractive strength of the potential energy. $\rho_N$ is the system's density ($\rho_N=N/L^2D$), where $N$ specifies the total number of particles ($N_A+N_B$). The preferential attraction towards a particular species is supervised via $\delta_\alpha$ for $\alpha \in$ ($A,B$). Substituting $\delta_B=0$ and $\delta_A=1$ produces wetting by $A$-type atoms. Further, the variable $z^{\prime}$ in Eq.~\eqref{eq1:wall} is defined as $z^\prime = z + \sigma/2$,  so that any singularity in the potential lies outside the simulation box. The wall at $z=0$ prefers $A$ (\ie, $\delta_A=1,\delta_B=0$), whereas the wall at $z=D$ is neutral (\ie, $\delta_A=0,\delta_B=0$) to both atom species. 

The particle-particle interaction is governed via standard pairwise LJ $12$-$6$ potential and has the form given as:
\begin{equation}\label{eq2:LJ}
\phi(r) = 4\epsilon_{\alpha\beta}\left[ \left(\frac{\sigma_{\alpha\beta}}{r}\right)^{12}- \left(\frac{\sigma_{\alpha\beta}}{r}\right)^6\right] .
\end{equation} 
The parameter $\epsilon_{\alpha\beta}$ sets the energy scale while $\sigma_{\alpha\beta}$ is the overlap distance for $\alpha,\beta \in (A,B)$.  We implement a symmetric mixture by keeping $\epsilon_{AA}=\epsilon_{BB}=2\epsilon_{AB}=\epsilon$, $\sigma_{AA}=\sigma_{BB}=\sigma_{AB}=\sigma$, and $m_A=m_B=m$. The equilibrium phase behavior of such symmetric mixtures is documented in detail. \cite{SJK03,SJK06,SMJ06}Also, the potential is truncated at distance $R_c=2.5~\sigma$ \cite{MD17} and is also shifted to remove discontinuity in force at the cut-off. Hence, both quantities go smoothly to zero at the cut-off $R_c$. Further, for the ease of computation, we set $\epsilon$, $\sigma$, and $m$ to $1$, and the reduced time of $t_0=\sqrt{m\sigma^2/\epsilon} = 1$. 

Next, we discuss the protocol implemented in the simulation. We start with $N$ particles randomly distributed in the simulation box, and only one type (\emph{e.g.}, only $B$-type) assigned to them. For $t<0$, the system is heated at a high temperature $T=3$, long enough to remove any meaningful spatial correlations and to achieve a thermodynamic equilibrium. Also,  with no types assigned to particles for $t<0$ guaranteed no phase-separation and wetting. At $t=0$, the required fraction of particles are randomly assigned $A$-type, and the rest are kept $B$-type. Simultaneously, the system's thermostat temperature is fixed at $T=1\approx 0.7\,T_c$ (bulk $ T_c\approx 1.423$), \cite{SJK06, SMJ06} along with the introduction of walls. No wall for $t<0$ is to avoid any inhomogeneities arising in the mixture in the form of layers.

 The data collection starts for $t>0$ as the system proceeds to a new equilibrium state. We use the Nos\'e-Hoover thermostat to control temperature, known for its ability to preserve hydrodynamics. \cite{SSS10,SSS12,allen96} Newtons equations of motion are integrated numerically using the velocity-Verlet algorithm, \cite{allen96} with a time step $\Delta t = 0.01$ in LJ units.
 \subsection{Phenomenological Model (Diffusive Transport Only)}
We briefly describe here the numerical model of Puri-Binder to study SDSD. \cite{puri05,SK92,SK94,KSS10} The PB model is a CHC equation supplemented by two boundary conditions and a surface potential.\cite{SK94,SK92} We implemented the surface potential as flat $V(z)=-h_1$ in a microscopic thickness for $0\le z< 1$ and then decaying in a power-law fashion $V(z) =\,-h_1/z^{n}$ for $z\ge1$, where $h_1$ measures the strength of the potential and $n$ denotes its range. Then, the dimensionless evolutionary equation from the PB model with surface potential has the form 
\begin{equation}\label{eq3:CHC}
\begin{split}
\frac{\partial \psi(\vec{r},t)}{\partial t} = &\vec{\nabla}\cdot \bigg[ \vec{\nabla} \bigg\{ -\,\psi +\,\psi^3 - \,\frac{1}{2}\nabla^2 \psi -V(z) \bigg\} + \vec{\theta}(\vec{r},t) \bigg],\\
&z>0.\\
\end{split}
\end{equation}  
 The corresponding two dimensionless boundary conditions are as follows:
 \begin{equation}\label{eq4:bc1}
 \begin{split}
 \tau_0 \frac{\partial \psi (\vec{\rho},0,t)}{\partial t} = &\, h_1(\vec{\rho})\,+\,g\psi (\vec{\rho},0,t)\,+\,\gamma\frac{\partial \psi}{\partial z}\bigg|_{z=0}\, \\
 &+\,\tilde{\gamma}\nabla_{\rho}^2 \psi(\vec{\rho},0,t), \\
 &
 \end{split}
 \end{equation}
 
 \begin{equation}\label{eq5:bc2}
 	\begin{split}
 	0=\bigg[ \frac{\partial}{\partial z}\bigg{\{}  -\,\psi +\,\psi^3 - \,\frac{1}{2}\nabla^2 \psi -V(z) \bigg{\}} + \theta_z \bigg]_{z=0}.
 	\end{split}
 \end{equation}

The details of the model are provided in Ref.~\cite{puri05} and are briefly discussed here for completeness. The order parameter, $\psi(\vec{r},t)$, is the density difference of $A$ and $B$-type particles at a spatial position $\vec{r}$. It is defined as $\rho_A(\vec{r},t)\,-\,\rho_B(\vec{r},t)$, where $\rho_A(\vec{r},t)$ and $\rho_B(\vec{r},t)$, respectively, are densities of $A$ and $B$ at $\vec{r}$. The variable $\tau_0$ in Eq.~\eqref{eq4:bc1} sets the relaxational timescale of non-conserved kinetics at the surface through the first boundary condition. We set this variable to zero and replaced $\tau_0 \frac{\partial \psi (\vec{\rho},0,t)}{\partial t}$ with its static counterpart $\partial \psi / \partial t = 0$. The position $\vec{r}$ is further decomposed into $\vec{\rho}$ and $z$ to implement the surface potential in the $z$-direction. Also, we ignore the lateral diffusion [$\tilde{\gamma}\nabla_{\rho}^2 \psi(\vec{\rho},0,t) = 0$]  because of the homogenizing effect brought by the potential $V(z)$.  Equation~\ref{eq5:bc2} sets the $z$-component of the current at the surface to zero, as there is no flux across this surface. The quantities $h_1(\vec{\rho})$, $g$, $\gamma$, and $\tilde{\gamma}$ determine the equilibrium wetting phase diagram of the system. \cite{SK92,SK94}

	\section{\label{sec:level3}RESULTS AND DISCUSSION}
	\subsection{\label{sec:level3p1}PHENOMENOLOGICAL THEORY}
	
	Before we present our MD results, we recapitulate the relevant results associated with domain coarsening from the phenomenological theory developed by Puri \emph{et al.} \cite{SK01, puri05, SK02pre} for the case with diffusive transport. Though our MD simulations naturally incorporate hydrodynamic effects, we expect the PB results to still apply in the early time regime before bulk phase separation sets in. An important consideration is the depth profiles, where the gradient signifies the current $J_z$ in the $z$-direction and could be written in the form of  
	\begin{equation} \label{eq3:J}
	J_z \simeq -\,\frac{dV(z)}{dz}{\bigg\rvert}_{z=R_1} - \frac{\gamma}{Lh}.
	\end{equation} 
	
		Here, the first term is the contribution from the external force, operative in the form of a potential $V(z)~=~- \epsilon_{a}/z^{n}$. The second term produces the current due to a chemical potential gradient across the flat wetting layer and other morphological domains formed in bulk, separated by the depletion layer width $h(t)=R_2 (t) - R_1(t)$ [where $R_1(t)$ and $R_2(t)$ are the first and second zero-crossings in the depth profiles at time $t$]. The variable $h(t)$ is further approximated as $h(t) \simeq [(1-\psi_{0})/(1+\psi_{0})]R_1(t)$ after assuming that the bilayer of wetting and depletion maintain the overall initial composition $\psi_{0}$. 
 
When substituting $h(t)$ and a long-ranged power-law potential of form $V(z) = - \epsilon_{a}/z^{n}$ in Eq.~\eqref{eq3:J}, we get
\begin{equation} \label{eq7:Jz}
J_z \simeq -\frac{n\,\epsilon_a}{R_1^{n+1}} - \frac{\gamma}{LR_1}\left(\frac{1+\psi_0}{1-\psi_0} \right).
\end{equation}
Moreover, as $J_z = -dR_1/dt$, we finally get the following growth regimes \cite{puri05, SK01},
\begin{equation}\label{eq8:R1}
R_1(t) \sim
\begin{dcases}
(\epsilon_a ~ t)^{1/(n+2)}, & t \ll t_c \, ,\\
\sqrt{\frac{(1+\psi_0)}{f(\psi_0)(1-\psi_0)}} ~ (\gamma ~ t)^{1/3}, & t \gg t_c\, .
\end{dcases}
\end{equation}  
where, the definition of $f(\psi_0)$ corresponds to the bulk length scale growth $l(t)=f(\psi_0)(\gamma t)^{1/3}$, and its expression can be found in Ref. \cite{SK01} The crossover time $t_c$ can be obtained by equating the early-time and late-time length scales as (for $n>1$ )

\begin{equation} \label{eq:t_c}
	t_c \sim h_1^{3/(n-1)}\;\gamma^{-(n+2)/(n-1)} \;\Bigg[\frac{f(\psi_0)(1-\psi_0)}{(1+\psi_0)}\Bigg]^{3(n+2)/[2(n-1)]}.
\end{equation}

 From the above equation, it is evident that the potential-dependent regime [$R_1(t) \sim t^{1/(n+2)}$] depends on the surface field strength $\epsilon_a$ as well as the exponent $n$, but is independent of the composition ratio of the system.

	\subsection{\label{sec:level3p2}QUALITATIVE AND QUANTITATIVE ANALYSIS}

		The simulation results presented here correspond to a temperature quench to $1 \approx 0.7T_c$ $( \text{bulk } T_c \approx 1.423)$ \cite{SJK06,SMJ06}  while having different compositions of wetting and non-wetting components. The wetting parameters are $\epsilon_r=0.6$ and $\epsilon_a=0.5$, which result in CW equilibrium morphology.\cite{SK10}  We primarily investigated a set of four different off-critical compositions, with two each belonging to wetting by the majority ($\psi_{0} \in \{0.6,0.4\}$) and by the minority ($\psi_{0} \in \{-0.4,-0.6\}$).
	
\begin{figure}[!htb]
		\centering
		\includegraphics[width=\linewidth]{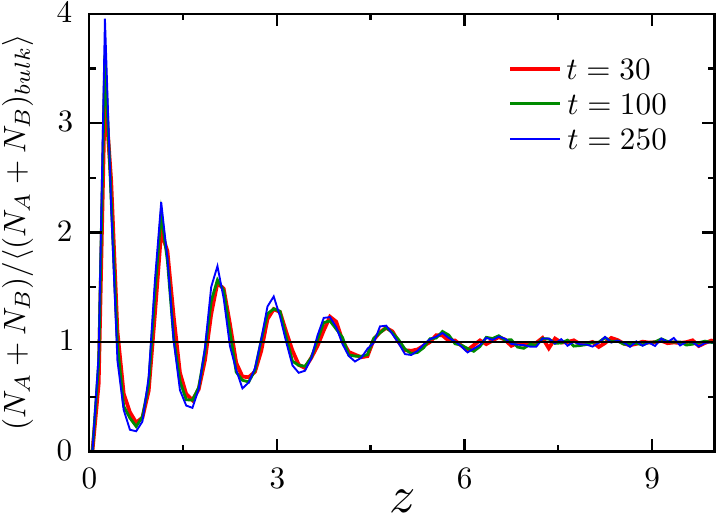}
		\caption{The laterally-averaged fluid number density is plotted along the $z$-direction for the overall composition of $\psi_{0}=-0.4$. $\langle(N_A+N_B)_{bulk}\rangle$ is the total number of particles under the homogeneous conditions for a layer inside the bulk. The black horizontal line represents system's number density $\rho_N$. The data is shown for three different times specified therein and shows no substantial variation during the simulation. The density fluctuations for other compositions are similar and not presented here for simplicity. }
		\label{fig:figure0}
	\end{figure}

		The flat walls for $t>0$ induce an oscillating fluid density in the vicinity of the walls, with the formation of layers of particles aligned perpendicular to the walls for some distance ($\approx 7$ $\sigma$). This layering effect is attributed to the smoothness of the flat walls and has been seen earlier. \cite{PWA94prl, PWA93} The oscillatory fluid density near the walls is presented in Fig.~\ref{fig:figure0} for a single composition of $\psi_{0}=-0.4$ (rest of the compositions show similar density fluctuation and are not shown here). The microscopic scale of density oscillations near the wetting wall affects the quantification of the wetting layer thickness, $R_1 (t)$. The measurement of $R_1(t)$ requires coarsening the simulation box into the slabs of size $L_x \times L_y \times \Delta z$, where $\Delta z$ is the slab's thickness in the $z$-direction. We then compute the laterally-averaged order-parameter $\psi_{\text{av}}(z,t)$, defined as \begin{equation}\label{eq5:psiav}
		\psi_{\text{av}} (z,t) = \frac{1}{L_x \times L_y} \int_{0}^{L_x} \int_{0}^{L_y} \psi(x,y,z,t) \;dx\;dy,
		\end{equation}  in all these slabs at different depths $z$ from the wetting surface and at time $t$. $R_1(t)$ is marked by a distance where the transition from an $A$-rich to a $B$-rich slab occurs for the first time along the $z$-direction. Therefore, if one chooses to extract $R_1(t)$ with microscopic precision, which is necessary to extract early time dynamics of the wetting layer, then the data is masked by the noise from the non-uniform distribution of particles for the slabs near the surface. For $\Delta z < \sigma$, the actual wetting layer growth is overshadowed by the jumps of the $R_1(t)$ between the slabs exhibiting the peaks of the density profiles. However, if we opt for a scale larger or equal to $\sigma$, we lose precision in the measurement of $R_1(t)$. Moreover, we compute laterally-averaged fluid number density $(N_A + N_B)/\langle(N_A+N_B)_{bulk}\rangle$ in each slab and present it in Fig.~\ref{fig:figure0}. We chose a slab width ($\Delta z$) of $0.1\sigma$. Here, $N_A$ and $N_B$ are the number of $A$- and $B$-type particles in the slab and $\langle(N_A+N_B)_{bulk}\rangle = \rho_N \times L_x \times L_y \times \Delta z$.

	\subsubsection{\label{sec:level3p1p1} Domain morphology in systems for majority and minority wetting}

		\begin{figure}[!htb]
		\centering
		\includegraphics[width=\linewidth]{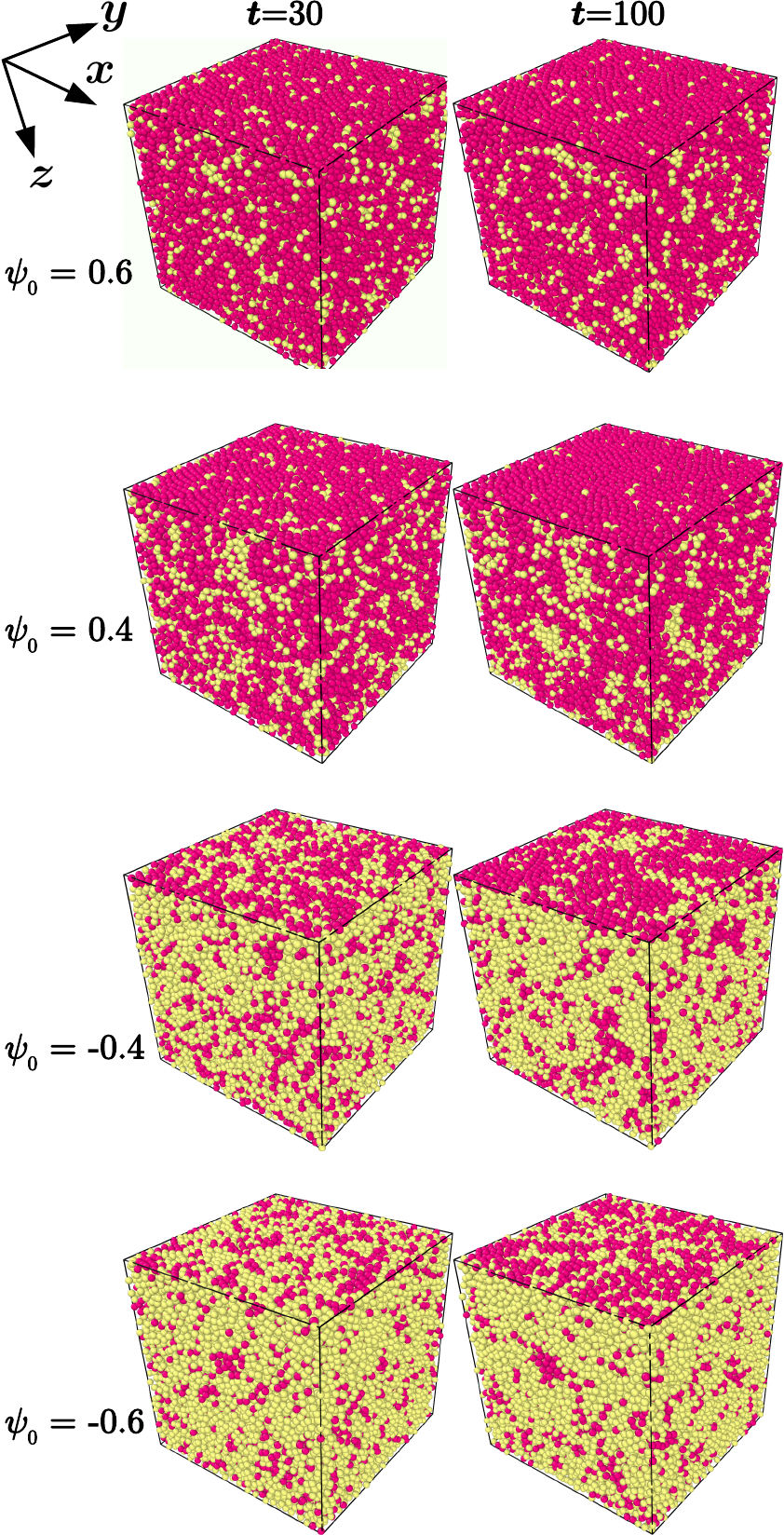}
		\caption{Pattern evolution in a binary mixture undergoing SDSD. The system is contained within a cubic box of size $L_x=L_y=L_z=L=32$, with the walls at $z=0$ (top) and at $z=32$ (bottom). Different rows and columns correspond to different composition ratios of wetting ($A$) and non-wetting ($B$) components for two different times, respectively. The wall at $z=0$ attracts $A$ (shown in pink) and gets completely wet by it. $B$-type particles are shown in yellow. The details of the simulation are in section~\ref{sec:level2}.  
		}
		\label{fig:figure1}
	\end{figure}
	
		We commence by showing the typical snapshots for $t>0$ obtained for different compositions. We present the combined images in Fig.~\ref{fig:figure1}, each row representing individual compositions. Since this study is focused only on the early time regime ($t<100$), we show the images for very early times where there is no substantial bulk phase separation. Although there is no relevant phase separation in the bulk, the wetting mechanism is operational. We see that for all the compositions, the wettable wall, at $z=0$, in time gets coated with the preferred particles, with some disparity present across the compositions due to the amount of the wetting component available. This disparity is central to the theme of this article.
	
		\begin{figure}[!htb]
		\centering
		\includegraphics[width=\linewidth]{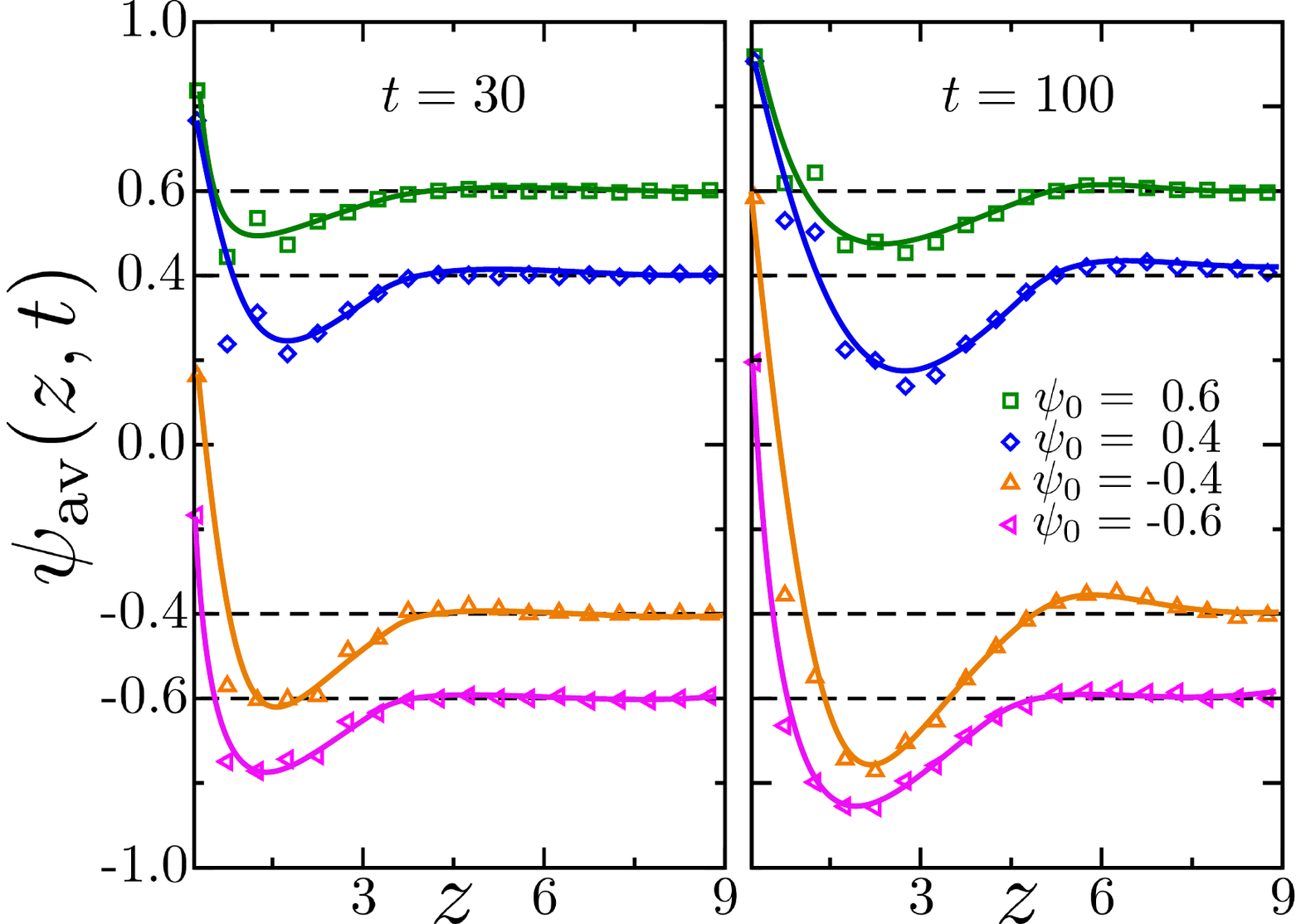}
		\caption{Laterally-averaged order parameter profile $\psi_{\text{av}}(z,t)$ for the representative evolution shown in Fig.~\ref{fig:figure1} at $t=30$ and $t=100$. The data and the dashed horizontal lines refer to different global compositions of $\psi_0 = 0.6,0.4, -0.4$ and $-0.6$. The solid lines are a guide to the eyes.}
		\label{fig:figure2}
	\end{figure} 
	
	We next present the laterally-averaged order parameter $\psi_{\text{av}} (z,t)$, with the definition in Eq.~\eqref{eq5:psiav}, which is a quantitative measure in most numerical and phenomenological studies addressing similar problems. \cite{puri05} Its experimental parallel is the depth profiling technique, and one may refer to early experiments by Jones \emph{et al.} for insights. \cite{RLE91,GCE93} We plot $\psi_{\text{av}}(z,t)$ in Fig.~\ref{fig:figure2} for four different compositions ($\psi_{0}=0.6,\;0.4,\; -0.4,$ and $-0.6$), where the dashed horizontal lines represent the overall order-parameter value for the systems, $\psi_{0}$. They act as a level about which $\psi_{\text{av}}(z,t)$ fluctuates and depicts the formation of lateral structures of the wetting and the depletion layer. Through time $t>0$, we notice the development of SDSD profiles, oscillating about their levels, with diminishing amplitudes, as they stretch into bulk. For the time shown here, the SDSD profiles exhibit a bilayer morphology with an initial wetting layer followed by a depletion layer. The formation of the bilayer morphology begins with the accumulation of preferred/wetting particles ($A$-type) above the attractive wall and the upwelling of the non-wetting ($B$-type) particles from the fresh wetting layer beneath. The exclusion of non-wetting particles from the wetting layer and the movement of wetting particles towards the wetting layer creates a depletion layer, separating the wetting layer from the bulk. Due to the conserved global concentration $\psi_{0}$ of the system with fixed $N_A$, the exchanges of the wetting particles ($A$) between the wetting layer and the rest of the system become dependent on its net amount $N_A$. Furthermore, the constraint is to achieve the equilibrium value of the $\psi_{\text{av}}(z,t)$ in the wetting layer. It is conspicuous from the plot, that the difference arising in majority versus minority wetting is from this constraint of maintaining the equilibrium value of $\psi_{\text{av}}(z,t) \simeq 1$, in the wetting layer, amid the abundance/shortage of wetting particles. The highly asymmetric nature of fluctuations about the $\psi_{0}$ in minority wetting causes a highly $B$-rich depletion layer. The composition of the depletion layer seems to be very off ($ \psi_0 \ll 0$ ) and could lie near the spinodal line. Such highly off-critical composition in the depletion layer creates a local barrier for the current responsible for the wetting layer growth. In contrast, for the majority wetting, the depletion layer always exhibits compositions closer to the critical compositions which offer weak resistance to the current. To sum it up, the compositions in the depletion layer shift toward the critical composition for the majority wetting, whereas they move away from criticality for the minority wetting.

	However, this composition dependence is never highlighted in earlier studies regarding the SDSD for such early times, irrespective of the presence of hydrodynamics and prevailing morphology in bulk. Other noticeable differences in the $\psi_{\text{av}}(z,t)$ profiles for different compositions are the gradients near the surface, as visible in Fig.~\ref{fig:figure2}. The profiles for minority wetting display sharp gradients at the boundary of wetting and the depletion layer, whereas $\psi_{\text{av}}(z,t)$ for the majority wetting with a surplus of wetting components varies slowly across the wetting layer. Overall, the composition differences create differences in the SDSD profiles and will additionally affect the dynamics because of the availability of wetting components. But to what extent in time and length scale these differences appear is what we attempt to characterize in the following discussion.

		\subsubsection{\label{sec:level3p1p2} The growth of the wetting layer}
		
		\subsubsection*{The early-time regime: Potential-dependent growth}
		
\begin{figure}[!htb]
			\centering
			\subfloat{\includegraphics[width=\linewidth]{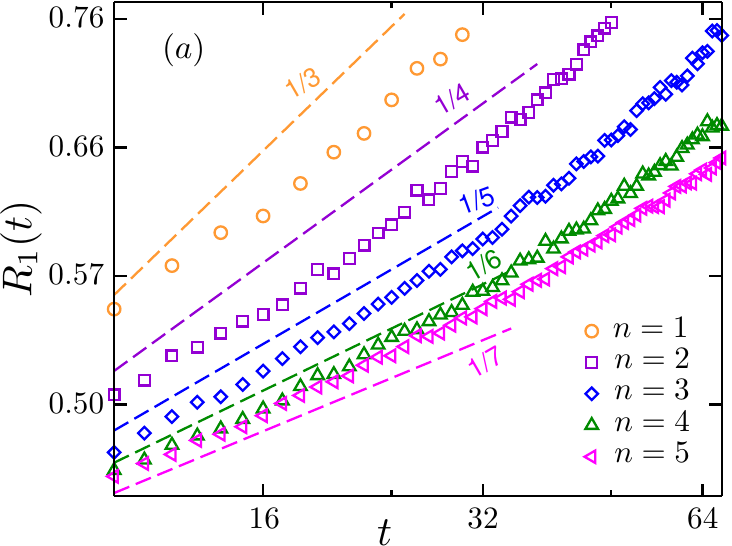}\label{figure3a}}
			
			\subfloat{\includegraphics[width=\linewidth]{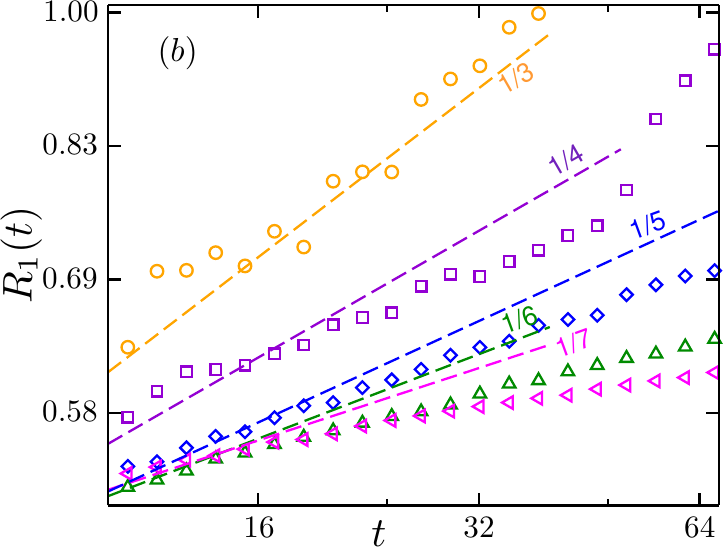}\label{figure3b}}
			\caption{Time dependence of wetting layer growth, $R_{1}(t)$, on a log-log scale for two composition ratios of $(a)$ $70:30$ and $(b)$ $30:70$. The attractive part of the $u_w (z)$ is $  -\epsilon_{a}/z^{n}$ where $n$ denotes its range and $\epsilon_{a}$ the interaction strength. $R_1(t)$ is extracted as \emph{first-zero crossing} from the SDSD profiles shown in Fig.~\ref{fig:figure2}. Phenomenological prediction for the time dependence of $R_{1}(t)$ for power-law potentials $V(z)$ is $\sim t^{1/(n+2)}$. The dashed lines with slopes of $1/3$, $1/4$, $1/5$, $1/6$, and $1/7$, correspond to the prediction for $n =1,2,3,4$ and $5$, respectively.}
			\label{fig:figure3}
		\end{figure}

	Next, we present results for the wetting layer's thickness $R_1(t)$. We chose the wetting layer, which lies next to the attractive wall and is followed by the depletion layer. The wetting layer's thickness at time $t$ is calculated from the \emph{first-zero crossing} in the SDSD profiles. This crossing is marked by a point on the line of overall $\psi_{0}$ at which the profile first crosses it from above. We extract such points from the profiles shown in Fig.~\ref{fig:figure2} and plot them against time on a log-log scale in Fig.~\ref{fig:figure3}.
		
		\begin{figure}[!htb]
			\centering
			\subfloat{\includegraphics[width=\linewidth]{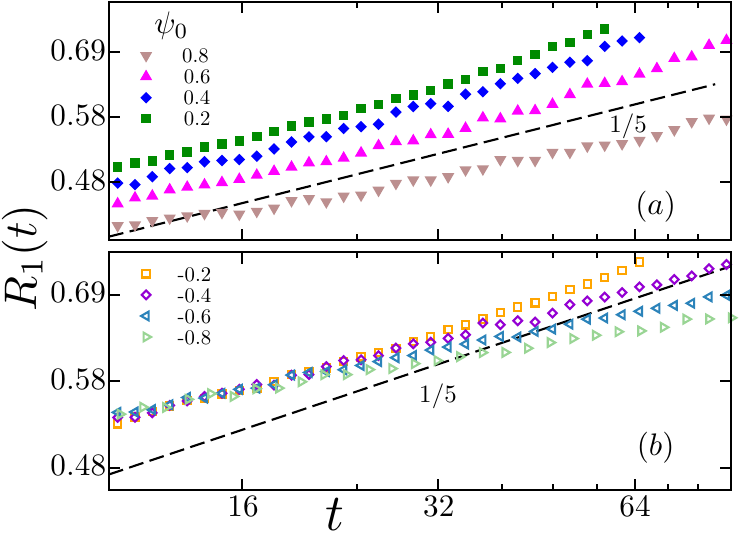}\label{figure4a}}
			
			\caption{Time dependence of $R_{1}(t)$ analogous to Fig.~\ref{fig:figure3} for early times . The data shown corresponds to different composition ratios for $n=3$ in $V(z) = -\epsilon_{a}/z^{n}$. The dashed lines have a slope of $1/5$, representing the potential-dependent growth ($R_1(t) \sim t^{1/(n+2)}$). The results are subdivided into $(a)$ the majority wetting, and $(b)$ the minority wetting.}
			\label{fig:figure4}
		\end{figure}
		
		We wish to reiterate that we focus on the dynamics corresponding to the very early time, which in previous studies belong to the potential dependent regime. Figure~\ref{fig:figure3}($a$) corresponds to the wetting layer growth for majority wetting ($\psi_{0}=0.4$) and Fig.~\ref{fig:figure3}($b$) for the minority wetting ($\psi_{0}=-0.4$). The different colors and symbols in the plot represent growth for different power-law potentials emanating from the lower wall at $z=0$. In the attractive part of employed potential given in Eq.~\ref{eq1:wall}, we model the interaction at various ranges, with $n=1$ behaving as a longer-range potential compared to $n=5$. For Fig.~\ref{fig:figure3}($a$), we get potential dependent growth right from the beginning of the simulation. It proves that the first term of the phenomenological current, $J_z$, in Eq.~\eqref{eq3:J} is active. Meanwhile, as stated earlier, the lack of any substantial phase separation during this time regime leads to an ineffective second term in $J_z$. Thus, we did not encounter any anomalous results, and the wetting growth seemed driven entirely by the potential gradient following the potential-dependent growth regime described in earlier studies.

		However, the growth behaves differently in the case of minority wetting. Fig.~\ref{fig:figure3}($b$) depicts such a scenario for $\psi_{0} = -0.4$. The wetting-layer dynamics are in agreement with the PB scenario for $n=1,2$. However, for $n > 2$, the growth does not follow Eq.~\eqref{eq7:Jz}.  We see that the growth for different potential ranges does not follow Eq.~\eqref{eq8:R1} for $t << t_c$ ( $t_c$ is the characteristic time marking the crossover from potential-dependent growth to diffusive growth given in Eq.~\ref{eq:t_c}). Furthermore, the growth is slowed for each $n$ value, with the slowest for the shortest range present ($n=5$). Therefore, it would not be wrong to say that none of the terms in the current, $J_z$, is contributing.

		Additionally, to extract any composition dependence of the $R_1(t)$, we plotted $R_1$ vs. $t$ on a log-log scale in Fig.~\ref{fig:figure4} for different compositions ( Fig.~\ref{fig:figure4}($a$) for $\psi_{0}>0$ and Fig.~\ref{fig:figure4}($b$) for $\psi_{0}<0$). It is done for $n=3$, which corresponds to non-retarded \emph{van der Waals} (\emph{vdW}) interactions. By looking at Fig.~\ref{fig:figure4}($a$), we realize that the power-law growth exponent in the majority wetting is composition-independent and follows the discussed phenomenology. However, the minority wetting growth in Fig~\ref{fig:figure4}($b$) is slower than the expected theoretical prediction and is even further subdued with the lowering of the wetting particles ($A$-type particles).
		 
			We understood that for a single time period, the wetting by majority fulfills the theoretical prediction, and the growth is powered by the first term of the phenomenological current, $J_z$ in Eq.~\eqref{eq3:J}. In contrast, the wetting layer growth plummets and the phenomenological current is no longer appropriate for the cases of minority wetting and/or when the potential range is shortened. Moreover, such disparity in the majority and minority wetting is also reflected in the results from the PB model when we numerically solve the Eqs.~\eqref{eq3:CHC},~\eqref{eq4:bc1}, and \eqref{eq5:bc2} (results not shown here). The matching results from the diffusion-limited PB model confirm the absence of any hydrodynamical effects.   
		
		After observing the different results for the minority component wetting ($\psi_{0}<0$), we look for an explanation. The appropriate description of the slower growth seemed beyond the ambit of phenomenological theory. However, the differences in the $\psi_{\text{av}}(z,t)$ profiles highlighted in Fig.~\ref{fig:figure2} are sufficient to address the growth disparity. As mentioned earlier, the asymmetric nature of the SDSD profiles in the cases of minority wetting de-accelerates the growth rates for the early-time regimes. The depletion layer has compositions close to nucleation regimes and therefore could not sustain adequate currents for the wetting layer. To reduce this asymmetry, we exercised with lower field strengths $\epsilon_a$ for different compositions. Lowering the field strengths results in lower equilibrium values of $\psi_{\text{av}}(z,t)$ in the wetting layer. Therefore, choosing a slightly lower field strength for minority wetting relaxes the constraint of achieving $\psi_{\text{av}}(z,t) \simeq 1$ in the wetting layers. The relaxations of equilibrium values affect the bilayer morphology of the wetting and depletion layer. Now, the composition fluctuations do not deviate too much from bulk values, and therefore the depletion layer has compositions not too close to the spinodal line.
		
		\begin{figure}[!htb]
			\centering
			\subfloat{\includegraphics[width=\linewidth]{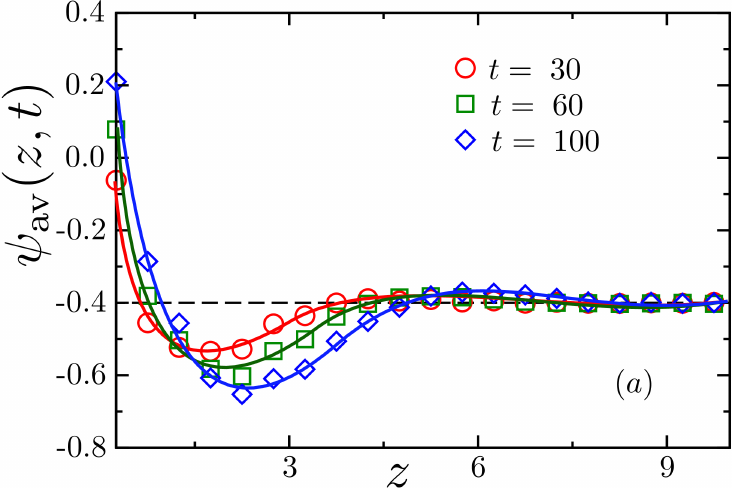}\label{figure8a}}

			\subfloat{\includegraphics[width=\linewidth]{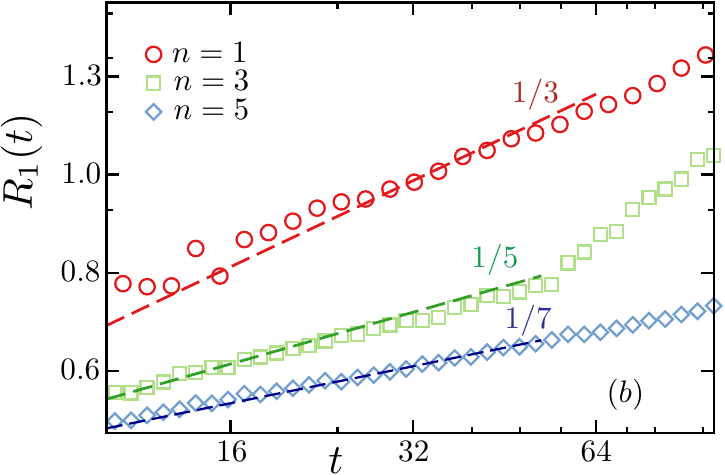}\label{figure8b}}
			\caption{$(a)$ Laterally-averaged order-parameter profiles analogous to Fig.~\ref{fig:figure2}, but for $\epsilon_a = 0.3$. The range of the potential is set to $n=3$. $(b)$ The time dependence of wetting layer growth for different potential ranges of $n=1,3,$ and $5$. The interaction strength, $\epsilon_a$, is equal to $0.3$ again. Dashed lines have slopes of $1/3$, $1/5$, and $1/7$.    }
			\label{fig:figure8}
		\end{figure}
	
	We present the results of minority wetting ($\psi_0=-0.4$) for wetting parameters of $\epsilon_a=0.3$ and potential range $n=3$ in Fig.~\ref{fig:figure8}. The results are analogous to what is shown in Fig.~\ref{fig:figure3}, but for weaker field strength. It is clear from the $\psi_{\text{av}}(z,t)$ profiles that the asymmetry appearing in the SDSD profiles for $\psi_0=-0.4$ in Fig.~\ref{fig:figure2} is brought down. Thus, the composition of the depletion layer remains inside the spinodal regime. The change is visible in the first minimum of the SDSD profiles when we compare the results from Fig.~\ref{fig:figure2} and Fig.~\ref{fig:figure8}($a$). The minimum values of $\psi_{\text{av}}(z,t)$ for $\epsilon_a=0.5$ shown in Fig.~\ref{fig:figure2} for $t=30$ and $100$ lie in between $-0.6$ and $-0.8$, whereas for $\epsilon_a=0.3$, there is a slight increase and the minimum values stay between $-0.5$ and $-0.7$. The resultant composition variation about the global composition is strictly guided by the conservation of order parameters along the SDSD profiles. Furthermore, in Fig.~\ref{fig:figure8}($b$), we plot the time dependence of the wetting layer for similar wetting parameters of Fig.~\ref{fig:figure8}($a$). It is clearly shown here that we recovered the potential-dependent regime with correct growth exponents from the Puri-Binder model for our minority wetting case with $\psi_0=-0.4$. 
		
			\begin{figure}[!htb]
			\centering
			\includegraphics[width=\linewidth]{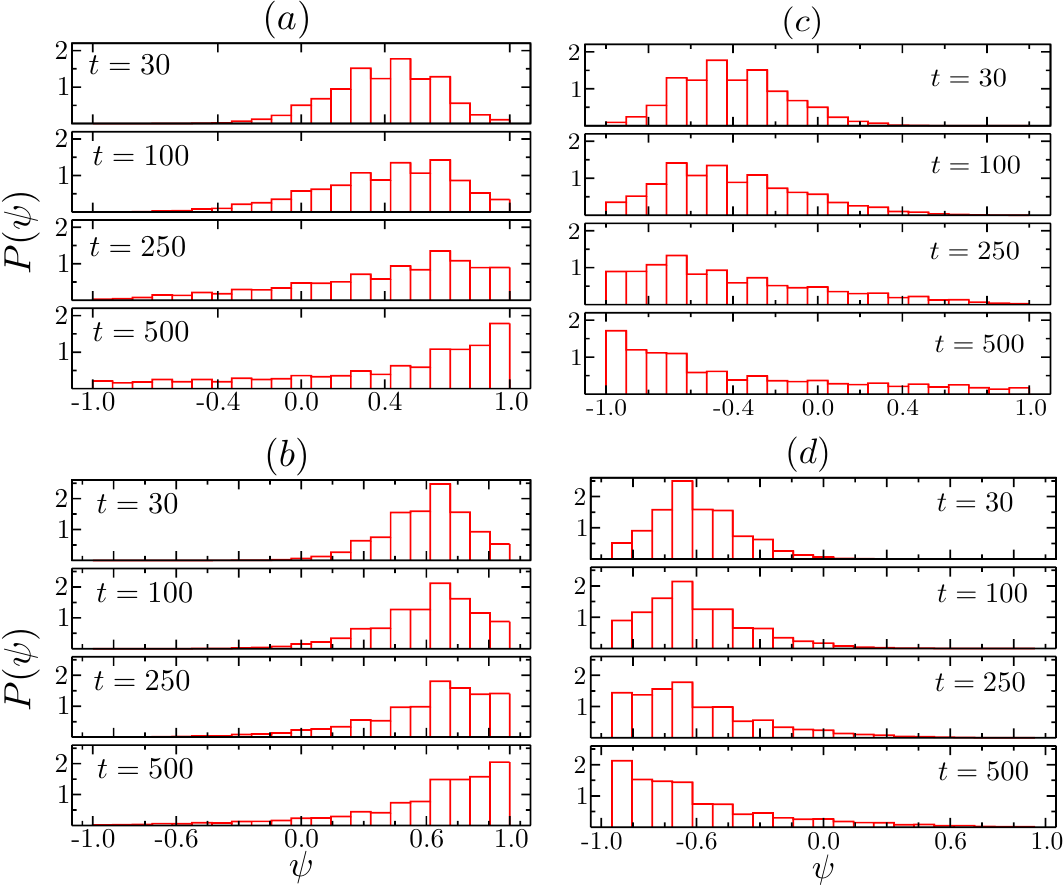}
			\caption{The probability distribution of the order-parameter values, $P(\psi)$, in phase-separating mixtures with different global composition ratios ($\psi_{0}$) and at different times. The data for respective compositions is arranged as such: $(a)$ $\psi_{0}=0.4$, $(b)$ $\psi_{0}=-0.4$, $(c)$ $\psi_{0}=0.6$, and $(d)$ $\psi_{0}=-0.6$, with the time specified alongside each distribution.}
			\label{fig:figure6}
		\end{figure}

			We have addressed the disparity in wetting layer growth across the majority and minority wetting. The mentioned resolution of lowering the asymmetry in the amplitude of fluctuations about the bulk value applies to both the rising off-criticality and the shortening of the range. In the latter, the asymmetry is accentuated by the increased field concentration at the wetting wall for the shorter-range potentials. Therefore, when one increases $n$ to decrease the potential range, the depletion layer widens and is shifted towards the spinodal line more, as the wetting layer accommodates more and more $A$-type particles quickly, showing steeper gradients in the SDSD profiles than shown in Figs.~\ref{fig:figure2} and \ref{fig:figure8}($a$). We have exercised with different $n$ and $\epsilon_a$ values and saw what we discussed here. However, to keep this work clean and concise, we are not presenting the results here. We plan to exhaustively investigate the role of field strength across majority and minority wetting in our next work.

			In Fig.~\ref{fig:figure6}, we plot the probability distribution of $\psi$ values computed after coarsening the simulation box in cubic unit cells of size $(2\sigma)^3$. The histogram of the data is computed for different $\psi_{0}$ and also at different times. Moreover, the data for the distributions is collected from $10$ different runs for each $\psi_0$. The motive behind such data presentation is to determine the degree of phase separation that has happened. This is important as the second term in the current, $J_z$ in Eq.~\eqref{eq3:J}, will be switched on once we get morphologies of relevant length scales. Moreover, the contribution of this second term could be negative in the case of majority wetting ($\psi_{0} >0$) due to the negative curvature of the domains above the wetting layer, which will deaccelerate the wetting layer growth. Alternately, the second term will provide positive feedback to the current for the minority case ($\psi_{0} < 0$). However, we notice that for the early-time study presented, there is no substantial phase separation, and the contribution of the second term to the current is absent or minimal.

	\section{\label{sec:level5}SUMMARY}
	Finally, we conclude by summarizing the significant results obtained from our molecular dynamics (MD) simulations. We simulated the ubiquitous process of surface-directed spinodal decomposition in a binary fluid mixture ($AB$) on a flat surface. The surface at $z=0$ is completely wet by $A$-type particles and the system exhibits the formation of structures of wetting and depletion layers perpendicular to the surface. The flatness of the wall also induces the layering of particles along the $z$-direction to some distance ($\sim 7\sigma$). The layering causes some problems in calculating the growth of the wetting layer when measured with microscopic precision in MD. Due to the layers of size $\sim 1\sigma$, the $z$-coordinates of the particles near both walls get discretized and manifest themselves as jumps in the growth of wetting layer thickness when measured with microscopic precision.   
	
	We focused only on the early-time dynamics of the wetting layer, as the asymptotic regime could not be reached due to the layering-induced effects of the system. Moreover, the extent of our study includes the effects of the amount of wetting particles present in the phase-separating mixture. We used the phenomenological theory to explain the underlying mechanism responsible for the observed growth laws. 
	
	We used phenomenological current from the Puri-Binder model and tested its applicability in early-time dynamics for different composition ratios. During this time regime, we do not notice any substantial phase-separation, and so the chemical-potential term in the current does not interfere with the growth.
	
	The wetting layer growth for the cases with a surplus of wetting particles followed the theoretical predictions, with the first term in the phenomenological current applicable. However, the wetting layer growth is slowed down and the first term is no longer adequate when the system lacks the wetting component. The slow growth is due to the constraint of the equilibrium order-parameter value of $\psi_{\text{av}}(z,t) \simeq 1$ in the wetting layer which shifts the depletion layer towards the nucleation regime. The depletion layer then forms a local barrier for the current responsible for the wetting layer growth. After relaxing that constraint ($\psi_{\text{av}}(z,t) < 1$) by lowering the field strength of attractive surface potential, we could recover the potential-dependent growth for a minority wetting case. The lower equilibrium value of $\psi_{\text{av}}(z,t)$ in the wetting layer damps the fluctuations in the bulk value. So, the depletion exhibits a lesser shift towards the nucleation regime.  A more systematic investigation is required for a better understanding of fluctuation growth for different values of the field strength and potential ranges. We consider this work as a move towards that goal.

	\begin{acknowledgments}
	P.K.J.  acknowledges the financial support from Science and Engineering Research Board (SERB), Department of Science and Technology (DST), India (Grant Nos. CRG/2022/006365, ITS/2024/001866) and  IIT Jodhpur for a Seed Grant (I/SEED/PKJ/20220016). S.P. is grateful to the Department of Science and Technology (DST), India for support via a J.C. Bose fellowship. 
	\end{acknowledgments}
	
	\vspace{5 mm}
		All authors declare they have no competing interests.	
		
	\section*{Data Availability}
	The article contains all the necessary information to reproduce the results presented.
\bibliographystyle{apsrev4-1}
\bibliography{offsdsd.bib}	
\end{document}